\documentstyle[prl,aps,twocolumn]{revtex}
\begin{document}
\draft

 {\bf Comment on "Spin relaxation in quantum Hall systems"}.

In the recent publication \cite{Apel} the authors have considered  the
spin relaxation in a 2D quantum Hall system for the filling factor $\nu \simeq 1$.
 The interest to  this
system is determined by the fact that the spin relaxation here is due to the
collective excitations, the so-called spin-excitons. Thus,  correct description
requires taking into account the Coulomb interaction and here we deal with the
delocalized states.
The authors  considered only one spin flip mechanism  among the three
possible \cite{Khaet}
spin-orbit related ones. This direct
spin-phonon coupling
is described by the
following term in the Hamiltonian
  \begin{equation}
\hat {\cal H}_{so}=\frac{1}{2} V_0  {\hat{\mbox{\boldmath $\sigma$}}}
\cdot {\hat {\mbox{\boldmath $\varphi$}}}; \/
 \hat \varphi_x = \frac{1}{2}\{u_{xy},\hat p_y \}_+,
\label{1}
\end{equation}
where $\hat {\bf p}$ is 2D electron momentum operator, the z-axis coincides with the normal
to the 2D plane, $\{,\}_+$ denotes the
anticommutator, $u_{ij}$ is the lattice strain tensor due to the acoustic
phonons, and  $V_0 = 8\cdot 10^7 cm/s$.
\cite{Dyak}
The authors came to the conclusion that the spin relaxation time due
to this mechanism is quite short:  around $10^{-10}$ s at B=10 T
(for GaAs) which is much shorter than the typical time ($10^{-5}$ s)
obtained in Ref.\cite{Frenkel} while considering the spin relaxation of 2D
electrons in a quantizing magnetic field without Coulomb interaction and for the
same spin-phonon coupling Eq(\ref{1}).
The authors \cite{Apel}  related this fact to the presence of the Coulomb interaction
and argued that now they are able to explain the earlier experimental data by
M.Dobers et al.
I show that their
conclusion about the value of the spin-flip time is wrong and have deduced the correct time
which is by several orders
of magnitude longer. I also discuss the
admixture mechanism \cite{Khaet} of the spin-orbit interaction.
The authors \cite{Apel} obtained the following expression for the spin-
relaxation rate:
 \begin{equation}
\frac{1}{\tau_{so}}=\sum_{{\bf Q},i=x,y} \frac{\pi}{2 \hbar}
 \mid \lambda^i({\bf Q}) \mid^2
\delta(E_{ex}(q) - \hbar \omega_{Q}),
\label{2}
\end{equation}
where ${\bf Q} = ({\bf q}, Q_z)$ and $\omega_{Q}$ are the phonon wave vector
and  the dispersion law, $E_{ex}(q)$ is the
2D dispersion law for the spin-excitons, {\mbox{\boldmath $\lambda({\bf Q}) $}}
is the  constant which couples the projected spin density of the electrons and
phonon creation operator in  the electron-phonon  part of the second-quantized
Hamiltonian. Using Eq.(\ref{1}), we get:
$ \mid \lambda^i({\bf Q}) \mid^2
= (\hbar^3 V_0^2/32 \rho s Q) q^2 q_i^2 \exp(-q^2 l_B^2/2)
 \mid \Lambda \mid^2$, where $\Lambda(Q_z)= \int dz \chi^2_0(z) \exp(iQ_z z) $,
 $s$ is the sound velocity, $ \rho $ the crystal mass density, $l_B$ the magnetic length
and $\chi_0 $ is the
 wave function in the z- direction.
Note that  parameter $x_0 = (\hbar s/\epsilon_c l_B)^2 \simeq ms^2/E_B
\approx 5\cdot 10^{-4}$ in GaAs, here $\epsilon_c=e^2/\kappa l_B, E_B$ are the Coulomb
 and  Bohr energies.
 Then the main contribution in Eq.(\ref{2}) comes from $
Q_z \simeq \epsilon_c/\hbar s \gg q \simeq 1/l_B$ and we obtain:
$1/\tau_{so} \simeq (0.48\sqrt{2}/4\pi^{3/2}) (V_0/s)^2 (\hbar/\rho l_B^5) x_0^{1/2}$.
The estimate for GaAs at B=10 T gives $\tau_{so} \approx 4\cdot 10^{-5} s$
(we assumed $\Lambda=1$ which can only overestimate the rate).  

\par
 It is much more essential, that with taking into account
the {\it spin-independent} interaction with the phonons
the energy relaxation time is much shorter than the spin relaxation time
and realistic situation corresponds to the spin relaxation of the 2D electron system
which has the lowest possible energy\cite{Dyck}.
Then the characteristic energy transferred to the phonon during the
spin-flip transition is much smaller than $\epsilon_c$ which appeared in
Eq.(\ref{2}). Now it is determined by Zeeman energy  $\Delta$ or temperature $T$.
(It remains unclear why the authors \cite{Apel} who assumed the temperature of the
electron system to be much smaller than $\Delta$, obtained  Eq.(\ref{2})).
Two different physical situations are possible \cite{Dyck}
depending on the degree of excitation of the spin system (i.e. the number of the spin-excitons
created in the 2D system). If this number is macroscopically
large (exceeds some critical value $N_c \propto T$), then the dominating channel of the spin
relaxation is the annihilation of two "zero" (i.e. with $q=0$) excitons  from
the condensate
with simultaneous generation of a "nonzero" exciton and phonon. The interaction  with
the phonons was spin-independent but the spin-orbit term in the Hamiltonian
$\beta (-\hat\sigma_x \hat p_x + \hat\sigma_y \hat p_y)$ which is due to the
absence of the inversion symmetry in the crystal cell leads to
an admixture of the state with the opposite spin and allows  spin-flip
 transition \cite{Khaet}. The nonexponential relaxation of the $S_z$ component
 is described by exactly the same equation as that obtained in Ref.\cite{Apel}
 but with  time (for the interaction with  piezo-phonons)
 $1/\tau_0 \simeq (\Delta Ms^2 M\beta^2/\hbar^3 \omega_c^2)
 ((eh_{14})^2 /\hbar s^3 \rho)$, which is written for $T, Ms^2 \ll \Delta$. Here $h_{14}$
 is the piezomodulus and $M$ the exciton mass. This time 
does not depend on the magnetic field and 
is  $10^{-5}\div 10^{-6} s$ depending on $\beta$. The mechanism described by Eq.(\ref{1}) gives
 a much smaller contribution to the rate. In the case of $N \ll N_c$ or near equilibrium 
 when the main process is a direct recombination of the "nonzero" excitons,
 the spin relaxation rate is proportional to the  temperature and can be also 
relatively small (time is longer than $10^{-6}$ s for $T< 1K$) \cite{Dyck}.

\bigskip \noindent
Alexander V. Khaetskii,

\par Institute of Microelectronics Technology, Russian Academy of Sciences,
142432, Chernogolovka, Russia

\bigskip \noindent


\end{document}